\documentclass[twocolumn]{aastex631}

\usepackage{amsmath}
\usepackage{multirow}
\usepackage{hyperref}

\newcommand\titlelowercase[1]{\texorpdfstring{\lowercase{#1}}{#1}}

\begin{document}

\title{The time evolution of $M_d/\dot M$ in protoplanetary disks as a way to disentangle between viscosity and MHD winds}

\author[0000-0003-2090-2928]{Alice Somigliana}
\affiliation{European Southern Observatory, Karl-Schwarzschild-Strasse 2, D-85748 Garching bei München, Germany}
\affiliation{Fakultat für Physik, Ludwig-Maximilians-Universität München, Scheinersts. 1, 81679 München, Germany}

\author[0000-0003-1859-3070]{Leonardo Testi}
\affiliation{Dipartimento di Fisica e Astronomia, Universita` di Bologna, Via Gobetti 93/2, 40122 Bologna, Italy}
\affiliation{INAF-Osservatorio Astrofisico di Arcetri, Largo E. Fermi 5, I-50125 Firenze, Italy}

\author[0000-0003-4853-5736]{Giovanni Rosotti}
\affiliation{Dipartimento di Fisica, Università degli Studi di Milano, Via Celoria 16, 20133 Milano, Italy}

\author[0000-0002-6958-4986]{Claudia Toci}
\affiliation{European Southern Observatory, Karl-Schwarzschild-Strasse 2, D-85748 Garching bei München, Germany}

\author[0000-0002-2357-7692]{Giuseppe Lodato}
\affiliation{Dipartimento di Fisica, Università degli Studi di Milano, Via Celoria 16, 20133 Milano, Italy}

\author[0000-0002-1103-3225]{Benoît Tabone}
\affiliation{Université Paris-Saclay, CNRS, Institut d'Astrophysique Spatiale, Orsay, France}

\author[0000-0003-3562-262X]{Carlo F. Manara}
\affiliation{European Southern Observatory, Karl-Schwarzschild-Strasse 2, D-85748 Garching bei München, Germany}

\author[0000-0003-3590-5814]{Marco Tazzari}
\affiliation{Institute of Astronomy, University of Cambridge, Madingley Road, CB3 0HA, Cambridge, UK}



\begin{abstract}

As the classic viscous paradigm for protoplanetary disk accretion is challenged by the observational evidence of low turbulence, the alternative scenario of MHD disk winds is being explored as potentially able to reproduce the same observed features traditionally explained with viscosity. Although the two models lead to different disk properties, none of them has been ruled out by observations - mainly due to instrumental limitations. In this work, we present a viable method to distinguish between the viscous and MHD framework based on the different evolution of the distribution in the disk mass ($M_{\mathrm{d}}$) - accretion rate ($\dot M$) plane of a disk population. With a synergy  of analytical calculations and 1D numerical simulations, performed with the population synthesis code \texttt{Diskpop}, we find that both mechanisms predict the spread of the observed ratio $M_{\mathrm{d}}/\dot M$ in a disk population to decrease over time; however, this effect is much less pronounced in MHD-dominated populations as compared to purely viscous populations. Furthermore, we demonstrate that this difference is detectable with the current observational facilities: we show that convolving the intrinsic spread with the observational uncertainties does not affect our result, as the observed spread in the MHD case remains significantly larger than in the viscous scenario. While the most recent data available show a better agreement with the wind model, ongoing and future efforts to obtain direct gas mass measurements with ALMA and ngVLA will cause a reassessment of this comparison in the near future.

\end{abstract}

\keywords{protoplanetary disks - accretion, accretion disks - planets and satellites: formation}


\section{Introduction}\label{sec:intro}

The gaseous component of protoplanetary disks has traditionally been described as undergoing viscous accretion (\citealt{LyndenBellPringle1974}, \citealt{Pringle1981}). In recent years, however, a growing observational evidence is challenging this picture, as the low levels of turbulence detected in protoplanetary disks appear incompatible with the observed evolution (\citealt{Pinte+2016}, \citealt{Flaherty+2018}, \citealt{Rosotti2023-Review}). The best alternative to the classic viscous scenario is currently provided by MHD disk winds, originally proposed by \cite{BlandfordAndPayne1982}. This model has gained increasing popularity in the recent years, as several studies (see \citealt{Lesur2021reviewMHD} for a review) have shown it to reproduce the key evolutionary features of protoplanetary disks; moreover, \cite{Tabone+2021a} have developed a simple analytical parametrization, making it a valid alternative to the viscous theory.

A compelling question is which of these mechanism, or which combination of the two, drives angular momentum transport in protoplanetary disks \citep{Manara+2022-PPVII}. Answering this question has proven to be a surprisingly difficult task: even though the two models are in principle well distinguishable through their characteristic theoretical predictions, the observational counterpart is lagging behind (e.g., \citealt{Rosotti+2019}, \citealt{Ilee+2022}). A good example of this problem is viscous spreading, a fundamental feature of viscous evolution that causes the gaseous component of disks to expand in radius as they evolve. As MHD evolution does not show a similar behavior \citep{Zagaria+2022b}, it would in principle be a good candidate for disentangling between the two predictions: however, the high sensitivity required to detect it has until now represented a limit. While Class 0 objects are widely accepted to be born small ($< 60$ au: \citealt{Maury2019}, also supported by the numerical experiments of, e.g., \citealt{Lebreuilly+2022}) and grow wider in the first 1-2 Myr of evolution \citep{Najita&Bergin2018}, whether the radius of Class II disks increases or decreases with time is widely debated. Dust continuum radii are observed to be shrinking with time (\citealt{Hendler2020}, \citealt{Zagaria+2022b}), as an effect of radial drift, while gas observations (\citealt{Ansdell+2018-radiigas}, \citealt{Sanchis2021}, \citealt{Toci+2021}, \citealt{Long+2022}) have covered too small of a sample at too low sensitivities to draw firm conclusions. The advent of ALMA Band 1 \citep{Carpenter+2020Band1} and the next-generation VLA (ngVLA, \citealt{Tobin+2018ngVLA}) in the near future will allow to perform surveys of protoplanetary disks at unprecedentedly long wavelengths, which will play a crucial role in determining the leading evolutionary mechanism. At the same time, finding novel approaches to tackle this problem is crucial to obtain significant results.

In this Letter, we suggest a new method to distinguish between the two models from the population perspective: through a joint theoretical and population synthesis approach, we investigate the time evolution of disks in the disk mass - accretion rate plane, proving it to be a good approach for our goal. This work is structured as follows: in Section \ref{sec:methods} we describe the evolutionary prescriptions that we adopt and we discuss their numerical implementation. In Section \ref{sec:results} we present our results and we compare them with the observations. Finally, in Section \ref{sec:discussion} we discuss the implications of these results and draw our conclusions.

\section{Theoretical model}\label{sec:methods}

\subsection{Secular evolution}\label{subsec:theo_models}

The simulations presented in this work have been carried out using the 1D Python population synthesis code \texttt{Diskpop}. For a detailed description of the code, as well as its public release, we refer to our upcoming paper (Somigliana et al. in prep; earlier implementations of the code, its basic assumptions and features have been described in \citealt{Rosotti+Letter2019}, \citealt{Rosotti+2019}, \citealt{Toci+2021}, \citealt{Somigliana+2022}). The viscous and MHD evolution are implemented following \cite{LyndenBellPringle1974} and \cite{Tabone+2021a} respectively. In this section we briefly present both models, referring to the original papers for a deeper discussion.

In the viscous case, we solve the classic evolution equation

\begin{equation}
    \label{eq:viscousevo}
    \frac{\partial \Sigma}{\partial t} = \frac{3}{R} \frac{\partial}{\partial R} \left( R^{1/2} \frac{\partial}{\partial R} (\alpha_{\mathrm{SS}} c_s H \Sigma R^{1/2}) \right);
\end{equation}

\noindent following the prescription by \cite{ShakuraSunyaev1973}, the viscosity $\nu$ is modeled as $\alpha_{\mathrm{SS}} c_s H$, where $\alpha_{\mathrm{SS}}$ is a dimensionless parameter, $c_s$ is the sound speed, and $H$ is the height of the disk. Furthermore, assuming the viscosity to be a power-law of the disk radius for ease of solving the equation, $\nu = \nu_c ( R/R_c)^{\gamma}$ (where $\nu_c = \nu(R = R_c)$ and $R_c$ is a scale radius), the analytical solution by \cite{LyndenBellPringle1974} holds.

In the MHD case instead \citep{Tabone+2021a}, the evolution equation is given by 

\begin{equation}
    \label{eq:mhd_evolution}
        \begin{split}
            \frac{\partial \Sigma}{\partial t} =  \frac{3}{R} \frac{\partial}{\partial R} \left( R^{1/2} \frac{\partial}{\partial R} (\alpha_{\mathrm{SS}} c_s H \Sigma R^{1/2}) \right) \\
            + \frac{3}{2R} \frac{\partial}{\partial R} \left( \frac{\alpha_{\mathrm{DW}} \Sigma {c_s}^2}{\Omega} \right) - \frac{3 \alpha_{\mathrm{DW}} \Sigma {c_s}^2}{4 (\lambda-1) R^2 \Omega},
        \end{split}
\end{equation}

\noindent where $\Omega$ is the keplerian orbital frequency, $\lambda$ is the magnetic lever arm parameter, and $\alpha_{\mathrm{DW}}$ is a magnetic equivalent of  $\alpha_{\mathrm{SS}}$. Equation \eqref{eq:mhd_evolution} is a generalization of Equation \eqref{eq:viscousevo} if the gas surface density evolves not only because of the viscous torque (first term on the RHS) but also because of the effects of MHD disk winds, which extract angular momentum and induce a mass loss (second and third term on the RHS respectively). Assuming that both $\lambda$ and $\alpha_{\mathrm{DW}}$ are constant across the disk, and that $\alpha_{\mathrm{DW}} \propto {\Sigma_c}^{-\omega}$ (where $\Sigma_c = \Sigma(R = R_c)$), Equation \eqref{eq:mhd_evolution} can be solved analytically (see \citealt{Tabone+2021a}).

\subsection{Isochrones}\label{subsec:isochrones}

Isochrones are defined as the curves described by a population of objects of the same age in a given plane. In the case of protoplanetary disks, isochrones in the $M_{\mathrm{d}} - \dot M$ plane have been the focus of recent studies (\citealt{Lodato2017}, \citealt{Somigliana+2020}). For viscously evolving disks \citep{Lodato2017}, the isochrone reads

\begin{equation}
    \label{eq:isochrone_viscous}
    \dot M = \frac{M_{\mathrm{d}}}{2 (2-\gamma) t} \left[ 1 - \left( \frac{M_{\mathrm{d}}}{M_0} \right)^{(2 - 2 \gamma)} \right];
\end{equation}

\noindent the only free parameter in Equation \eqref{eq:isochrone_viscous} is the initial disk mass $M_0$, which only sets the starting point of the isochrone. Nonetheless, at late stages (when $M_{\mathrm{d}} \ll M_0$) all disks in a population are bound to reach the same locus on the $M_{\mathrm{d}} - \dot M$ plane: while this happens at different ages for each disk, depending on its viscous timescale $t_{\nu} = {R_c}^2/(3 (2 - \gamma)^2 \nu_c)$, a fully evolved population ($t \to + \infty$) will necessarily sit on the theoretical isochrone of the corresponding age. 

For MHD disks, the isochrone is defined as \citep{Tabone+2021a}

\begin{equation}
    \label{eq:isochrone_mhd}
    \dot M = \frac{1}{\omega (1 + f_{\mathrm{M}, 0}) t} M_{\mathrm{d}} \left[  \left( \frac{M_{\mathrm{d}}}{M_0} \right)^{- \omega} - 1 \right];
\end{equation}

\noindent Equation \eqref{eq:isochrone_mhd} depends not only on $M_0$, but also on the equivalent of $t_{\nu}$ in the MHD winds case, the initial accretion timescale $t_{\mathrm{acc}, 0}$ , through $f_{\mathrm{M}, 0}$ (determined by the disk radius - see \citealt{Tabone+2021a} for details). The interpretation of the isochrones in the two models is therefore different: while the viscous curves for all disks in a population lie on top of each other (except at the early stages, when $M_{\mathrm{d}} \sim M_0$), MHD evolution never loses memory of the initial conditions. This is because, depending on whether we fix $M_0$ or $t_{\mathrm{acc}, 0}$, we can define two types of isochrones for an MHD population. As a result, disks with a different $M_0$ will occupy an \textit{area} of the $M_{\mathrm{d}} - \dot M$ plane, rather than sitting on a single curve, and this will be the case even for evolved populations - which means that it is not possible to use the isochrones to obtain age estimates for disk populations. Based on this argument, we investigate whether the evolution of a population of disks in the $M_{\mathrm{d}} - \dot M$ plane could carry tangible signatures of the evolutionary model.

\subsection{Population synthesis}\label{subsec:popsynth}

In this work we adopt a population synthesis approach, which consists of generating and evolving a synthetic population of protoplanetary disks via numerical methods. We employ the Python tool \texttt{Diskpop}, which we expanded from our previous work \citep{Somigliana+2022} to include MHD disk wind evolution. In this section, we present a brief outline of the workflow, referring to the upcoming code release for a detailed description of the methods and the implementation.

First, we generate $N \sim 100$ stars, whose masses $M_{\star}$ follow the \cite{Kroupa2001} initial mass function. We then assemble a Young Stellar Object (YSO) by assigning a disk to each star: to determine the initial mass and radius of said disk, we assume that the initial disk mass and accretion rate scale as power-laws of the stellar mass ($M_{\mathrm{d}} \propto {M_{\star}}^{\lambda_{\mathrm{m}}}$ and $\dot M \propto {M_{\star}}^{\lambda_{\mathrm{acc}}}$). In our previous work \citep{Somigliana+2022} we have demonstrated how $\lambda_{\mathrm{m}, 0} \in [0.7, 1.5]$ and $\lambda_{\mathrm{acc}, 0} \in [1.2, 2.1]$ can reproduce the slopes of observed correlations of disk properties with stellar mass at later ages; we refer to that paper for a detailed discussion. We determine $M_{\mathrm{d}}$ and $\dot M$ for each disk drawing from a log-normal distribution, centered in the mean value computed via the power-law correlations and with a width ($\sigma$) of choice; $R_{\mathrm{d}}$ is then derived from considerations on $\dot M$ (see \citealt{Somigliana+2022} for details). The other relevant quantities besides $M_{\star}$, $M_{\mathrm{d}}$ and $R_{\mathrm{d}}$ are fixed in our model: Table \ref{tab:parameters} shows the parameters that we used in the simulations presented in this work, based on the disc evolution studies of \cite{Lodato2017} and \cite{Tabone+2021a} for viscosity and MHD winds respectively. While a detailed study of the parameters space is outside of the scope of this work, we have tested two more combinations of parameters (shown by \cite{Tabone+2021b} to reproduce the Lupus star-forming region) and we found that our results are independent on the particular combination chosen. Once the population of YSOs is generated, it is evolved following the viscous or MHD prescription via a 1D implementation of the models described in Section \ref{subsec:theo_models}. Although \texttt{Diskpop} allows to numerically solve the evolution equations, in this work we have used the analytical solutions to Equation \eqref{eq:viscousevo} and \eqref{eq:mhd_evolution}; it is therefore important to note that our results depend on the assumptions needed to obtain such solutions (e.g., the power-law scaling of viscosity with the disk radius).

\begin{table*}
\begin{center}
\begin{tabular}{c c c c c c c c c c c} 
 \hline
 \textbf{Model} & \textbf{Distributions} & \textbf{IMF} &  $\lambda_{\mathrm{m}}$, $\lambda_{\mathrm{acc}}$ & $\sigma_{\mathrm{M}}$, $\sigma_{\mathrm{R}}$ & $H/R$ \textbf{at} $R = 1$ \textbf{au} & $\alpha_{\mathrm{SS}}$ & $\alpha_{\mathrm{DW}}$ & $\omega$, $\lambda$ & $<t_{\mathrm{acc}, 0}>$ \\ [0.5ex] 
 \hline\hline
 \textbf{Viscous} & \multirow{2}{*}{log-normal} & \multirow{2}{*}{\cite{Kroupa2001}} & \multirow{2}{*}{1.5, 2.1} & 1 dex & \multirow{2}{*}{0.03} & $10^{-3}$ & $0$ & $0$ & \multirow{2}{*}{0.8 Myr}\\
 [0.3ex]
 \textbf{MHD} & & & & 0.65 dex, 0.52 dex & & 0\footnote{Although the MHD model of \cite{Tabone+2021a} allows both $\alpha_{\mathrm{SS}}$ and $\alpha_{\mathrm{DW}}$ to be non-zero, Equation \eqref{eq:mhd_evolution} in the $\omega \neq 0$ case can only be solved analytically if $\alpha_{\mathrm{SS}} = 0$.} & $10^{-3}$ & 0.25, 3 & \\
 \hline
\end{tabular}
\caption{Parameters used in the viscous and MHD \texttt{Diskpop} simulations respectively. $\sigma_{\mathrm{M}}$ and $\sigma_{\mathrm{R}}$ are the width of the initial disk mass and radius, respectively. These values were chosen following the works of \cite{Lodato2017} and \cite{Tabone+2021a}.}
\label{tab:parameters}
\end{center}
\end{table*}

It is crucial to point out that disk dispersal is an intrinsic feature of MHD winds, but not of viscous evolution. Our code includes an observational effect by considering as dispersed disks with masses lower than $10^{-6} M_{\odot}$; this simulates a dispersal effect even in the viscous scenario, which would otherwise generate disks with infinite lifetime, that do not match the observed disk fraction (see Appendix \ref{appendix:df_and_af}). This problem is usually solved in the literature by adding other physical effects to the purely viscous model, such as internal photoevaporation (see e.g. \citealt{Hollenbach1994}, \citealt{ClarkeGendrinSoto2001}, \citealt{Owen+2011b}, \citealt{Picogna+2019}, \citealt{Emsenhuber+2023}). In order to account for the statistical effect of reducing our sample throughout the evolution caused by disk dispersal, we performed 100 simulations for both setups described in Table \ref{tab:parameters} and then considered not only the median evolution of the interesting quantities, but also the interval between the 25th and 75th percentile (see Section \ref{sec:results}).

\section{Results}\label{sec:results}

In this Section, we show the results of the evolution of viscous and MHD populations of protoplanetary disks in the $M_{\mathrm{d}} - \dot M$ plane: in particular, we consider the ratio of the two quantities (hereafter $t_{\mathrm{lt}}$, disk lifetime - see \citealt{Jones+2012}). We first discuss the expected evolution of the distribution of disk lifetimes from an analytical point of view (paragraph \ref{subsec:lifetimes_distr}), and then we confirm our theoretical results through \texttt{Diskpop} simulations (paragraph \ref{subsubsec:mean_and_width}); finally, we compare our results with the observations (paragraph \ref{subsec:comparison}).

\begin{figure}
    \centering
    \includegraphics[width = 0.49\textwidth]{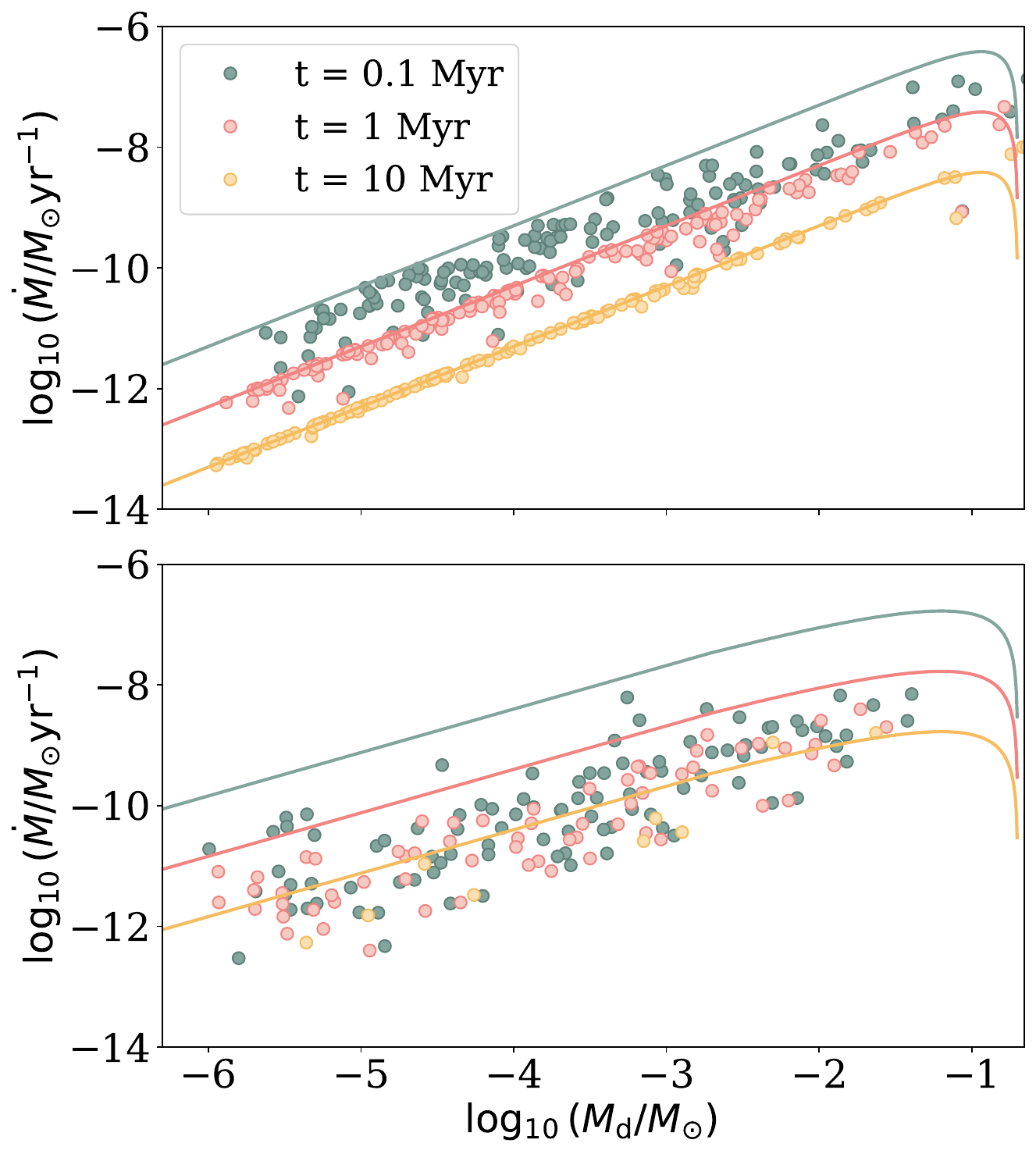}
    \caption{Time evolution of a synthetic population of disks, evolved via viscosity (top) or MHD winds (bottom), in the $M_{\mathrm{d}} - \dot M$ plane. The solid lines show the theoretical isochrones at ages 0.1, 1, and 10 Myr as per the legend, while the disks in the population at each age are represented by dots with the same color coding. While viscous disks tend to converge to the same isochrone at evolved stages, MHD disks show a larger dispersion.}
    \label{fig:isochrones_both}
\end{figure}

\subsection{Disk lifetimes distribution}\label{subsec:lifetimes_distr}

In the traditional viscous picture (\citealt{DullemondNattaTesti2006}, \citealt{Lodato2017}), disks lie on the theoretical isochrone (Equation \ref{eq:isochrone_viscous}) at a given age $t$ if their initial viscous timescale $t_{\nu, 0}$ is much shorter than $t$; as evolution proceeds, more and more disks reach this stage and therefore the population converges around the corresponding isochrone. As a consequence, the spread around the isochrones decreases with time: eventually, once the population is fully self-similar (i.e., its age is larger than all of the viscous timescales), the spread will be vanishingly small and the correlation between $M_{\mathrm{d}}$ and $\dot M$ will be perfectly linear. This trend is illustrated in the top panel of Figure \ref{fig:isochrones_both}: the solid lines show three theoretical isochrones at different ages, while the dots represent a synthetic population of $100$ disks obtained with \texttt{Diskpop} evolving in time with the same color coding. The aforementioned convergence to the theoretical isochrone starts as early as 1 Myr, while at 10 Myr the population is almost fully evolved and closely resembles the theoretical curve. From this argument, we can expect the moments of the distribution of $t_{\mathrm{lt}}$ to evolve in the viscous case as follows: (i) the mean value of $t_{\mathrm{lt}}$ will converge towards the actual age of the region, (ii) the spread will decrease until $t_{\nu} < t$ for every disk in the population, (iii) the skewness will increase. For a more detailed discussion on the expected and observed evolution of the skewness, we refer to Appendix \ref{appendix-skewness}.

The bottom panel of Figure \ref{fig:isochrones_both} shows a synthetic population of disks evolved via MHD winds in the $M_{\mathrm{d}} - \dot M$ plane. As discussed in paragraph \ref{subsec:isochrones}, the evolved population does not converge to the same isochrone: the large spread at all ages is such that making a prediction on the time evolution of the distribution of $t_{\mathrm{lt}}$ is not as straightforward as for a viscous population. \cite{Tabone+2021b} have shown that, assuming an exponential distribution of $t_{\mathrm{acc}, 0}$ (which is determined fitting the observed disk fraction), the distribution of $t_{\mathrm{lt}}$ does not depend on time; however, this result is specific of the exponential distribution. If we consider a different distribution of $t_{\mathrm{acc}, 0}$, that of $t_{\mathrm{lt}}$ for an evolved population may depend on time: this is the case for our choice of a log-normal distribution of $t_{\mathrm{acc}, 0}$, which can still reproduce both the disk and accretion fraction (see Appendix \ref{appendix:tlt}).

\subsection{Mean and width}\label{subsubsec:mean_and_width}

\begin{figure}
    \centering
    \includegraphics[width = 8cm]{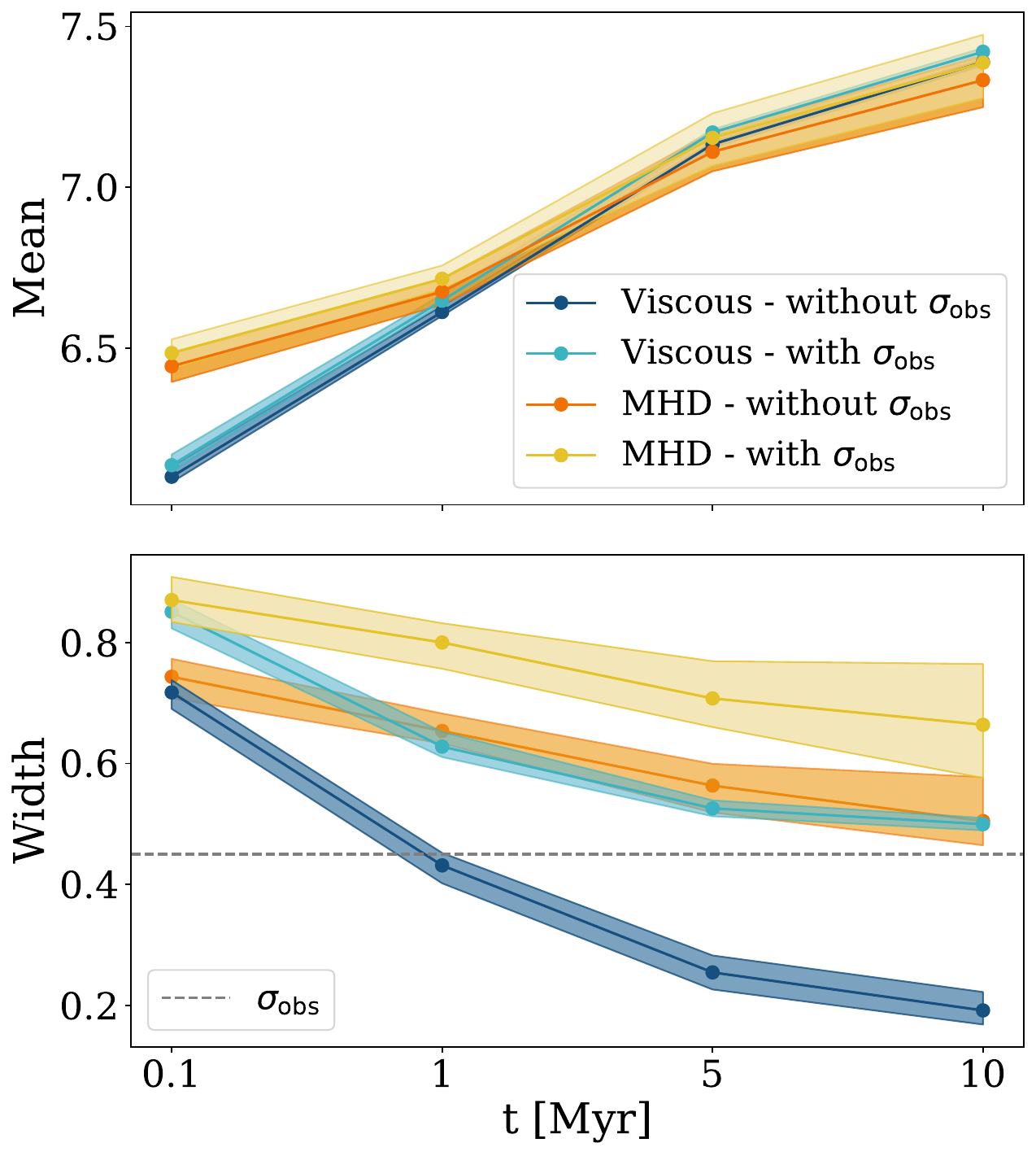}
    \caption{Time evolution of the mean (top) and width (bottom) of the distribution of $t_{\mathrm{lt}}$ for a synthetic population of protoplanetary disks. The solid lines represent the median values, while the shades cover the interval between the 25th and 75th percentile out of 100 simulations (to account for the statistical effect of disk removal). The blue and yellow lines refer to the viscous and MHD model respectively, with the lighter shades including the observational uncertainty. While the mean value of the distributions is not much affected by the presence of such uncertainty or the choice of the model, the spread shows quite some difference, exhibiting significantly higher values in the MHD than in the viscous case. The dashed line in the bottom panel marks the observational uncertainty.}
    \label{fig:mean_and_width}
\end{figure}

Figure \ref{fig:mean_and_width} shows the time evolution of the mean (top) and width (bottom) of the distributions of $t_{\mathrm{lt}}$ for the viscous (blue) and MHD (yellow) models. The lighter shades of both models include an additional observational uncertainty, $\sigma_{\mathrm{obs}}$, that we implemented by adding an extra spread on the disk mass and the accretion rate, of 0.1 dex and 0.45 dex respectively (as an estimate of the observational uncertainty, see \citealt{Manara+2022-PPVII}, \citealt{Testi+2022-Ofiucone}). As stated in Section \ref{sec:methods}, we performed 100 runs for each simulation: the solid line represents the median, while the shaded areas around it show the 25th-75th percentile intervals. As the MHD model removes disks more effectively, the sample size decreases more than in the viscous case, making the statistical fluctuations between different simulations larger: this leads the yellow lines to have broader shaded areas.

Considering the mean values of the distributions, adding $\sigma_{\mathrm{obs}}$ only slightly shifts the curves for both the viscous and MHD case, resulting in a negligible difference. The two evolutionary models differ at early stages ($< 1$ Myr), but soon reach a common behavior that makes them indistinguishable within the 25th-75th percentile intervals. On the other hand, the widths of the distribution (bottom panel) significantly differ from one case to the other. The viscous case without additional uncertainty (darker blue) steeply decreases, as expected from viscous theory \citep{Lodato2017} and discussed in paragraph \ref{subsec:lifetimes_distr}. This is not the case for the MHD prescription (orange): while the general trend is still decreasing, it is not as steep as the viscous, and ultimately does not tend to zero but rather to an evolved value determined by the initial conditions.

The convolution with observational uncertainty in the viscous case (light blue) significantly shifts the curve up, as well as modifying its shape. The total width of the distribution is the root sum squared of the intrinsic spread ($\sigma_{\mathrm{int}}$) and the observational uncertainty ($\sigma_{\mathrm{obs}}$), $\sigma_{\mathrm{tot}} = \sqrt{{\sigma_{\mathrm{int}}}^2 + {\sigma_{\mathrm{obs}}^2}}$. The intrinsic spread $\sigma_{\mathrm{int}}$, given by the initial conditions, tends to zero as discussed above: therefore, we expect the final width to tend to $\sigma_{\mathrm{obs}}$, which is exactly what we recover. This causes the evolved population to have a significantly larger spread than that predicted by theory. On the contrary, despite still being shifted at larger values as an effect of the additional uncertainty, in the MHD case (yellow) the shape of the curve is not dramatically modified. This is because $\sigma_{\mathrm{int}}$ is comparable to $\sigma_{\mathrm{obs}}$ at all times, which makes this argument strongly dependent on the initial condition: as the total spread is given by $\sqrt{{\sigma_{\mathrm{int}}}^2 + {\sigma_{\mathrm{obs}}}^2}$, the behavior of the MHD case will only be significantly different from the viscous case if $\sigma_{\mathrm{int}}$ is non negligible with respect to $\sigma_{\mathrm{obs}}$. In our previous work \citep{Somigliana+2022} we have shown how initial spreads of 0.65 dex and 0.52 dex for $M_{\mathrm{d}}$ and $R_{\mathrm{d}}$ respectively are able to reproduce the observed spreads around the correlations with the stellar masses; therefore, we set these values for the MHD simulation, while we choose a bigger spread of 1 dex for the viscous case, as it can better reproduce the observed values (see \ref{subsec:comparison}). 

As mentioned in Section \ref{subsec:popsynth}, the purely viscous model does not account for disk dispersal. Without exploring the whole parameter space, which is beyond the scope of this Letter, we have run a test model with photoevaporation, assuming the standard model of \cite{Owen+2010}, with a mass-loss rate of $10^{-10}$ M$_{\odot}$ yr$^{-1}$ following the latest constraints \citep{Alexander+2023}. The mean and the width of the distribution of $t_{\mathrm{lt}}$ increase with respect to the purely viscous case, but the difference is minimal and becomes negligible including the observational uncertainty; therefore, our conclusions are not affected.

\subsection{Comparison with the observations}\label{subsec:comparison}

\begin{figure}
    \centering
    \includegraphics[width = 8cm]{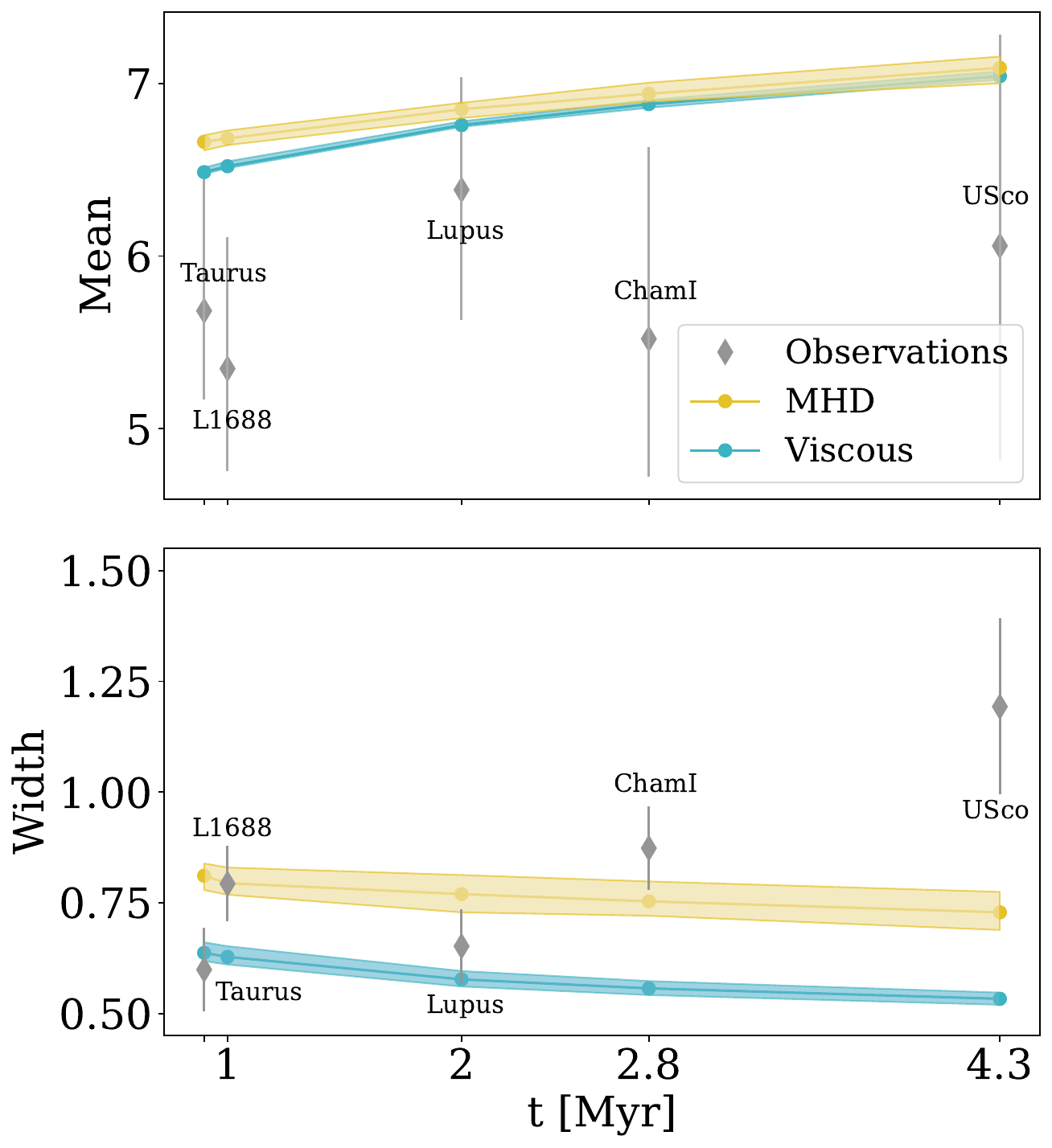}
    \caption{Comparison of the evolution of the mean (top) and width (bottom) for the viscous (blue) and MHD (yellow) models, including observational uncertainties, with the observations (gray diamonds). The shaded areas are as in Figure \ref{fig:mean_and_width}, while the gray bars represent the interval between the 16th and 84th percentiles (top) and the uncertainty on the width (bottom). While both models overestimate the mean values (see text for details), the evolution of the width of the distribution suggests a better match with the MHD model.}
    \label{fig:comparison_obs}
\end{figure}

In paragraph \ref{subsubsec:mean_and_width} we have shown the viscous and MHD predictions for the time evolution of the mean and width of the distribution of $t_{\mathrm{lt}}$; in this paragraph, we compare our results with observations of different star-forming regions. We used the table\footnote{The table is available at \href{http://ppvii.org/chapter/15/}{http://ppvii.org/chapter/15/}.} compiled by \cite{Manara+2022-PPVII} for Taurus, Lupus, Chameleon I and Upper Sco, and the data by \cite{Testi+2022-Ofiucone} for L1688 (to limit the contamination from sub-populations with different ages in the Ophiuchus complex).

Before commenting on the comparison itself, it is important to note that our simulations do not include dust evolution, making our definition of disk mass solely based on the \textit{gas} content of disks; on the other hand, the observed disk masses rely on sub-mm fluxes, tracing the \textit{dust} content instead. As the bulk of disk masses is in the gaseous phase, inferring the total mass from dust observations requires to i) constrain the dust-to-gas ratio in disks and ii) assume optically thin emission; however, as the accuracy of these assumptions is debated, the community is striving towards obtaining more reliable disk mass estimates (see \citealt{Bergin+2013HD}, \citealt{McClure+2016HD} for HD observations; \citealt{Veronesi+2021} for dynamical measurements; \citealt{Anderson+2022}, \citealt{Trapman+2022} for a combination of gaseous tracers). The results of the ALMA Large Programs AGE-PRO and DECO will further contribute to this goal; moreover, the advent of the ALMA Band 1 and ngVLA will allow to move to longer wavelengths, where dust emission is less optically thick \citep{Tazzari+2021}. In light of these forthcoming developments, our work can be considered a prediction that will be interpreted to its full potential with the results of this observational effort. The data comparison presented in the following is therefore intended as a state-of-the-art, which we anticipate to revise in the near future.

Figure \ref{fig:comparison_obs} shows the result of our comparison: the mean and width of the distribution are shown in the top and bottom panel respectively, and both include the viscous and MHD (blue and yellow line, as in Figure \ref{fig:mean_and_width} and \ref{fig:skewness_all}) numerical evolution. The gray diamonds represent the observed star-forming regions. None of the two evolutionary mechanisms reproduces the observed mean values, which are systematically lower. A potential reason for this mismatch could be an underestimation of disk masses; a difference of a factor as little as 3 in the observed masses would be sufficient to explain the discrepancy with the models - confirming the need to repeat this comparison with more accurate disk masses estimates. Moreover, \cite{Zagaria+2022binaries} have shown how taking stellar multiplicity into account can explain the high accretors in Upper Sco; we expect this effect to shift the theoretical prediction to lower values of $t_{\mathrm{lt}}$ for evolved populations. Dust growth and evolution prescriptions, which were not included in this work, are also likely to play a role as they can better explain the observed disk mass - accretion rate correlation \citep{Sellek+2020-dustyorigin}. The width of the distribution, on the other hand, provides more interesting results. The viscous prediction manages to marginally recover observed values at the earliest evolutionary stages, but as such values increase in time, the discrepancy with the viscous expectation grows larger and larger. This result was already anticipated by \cite{Manara+2020-UpperSco} (see also \citealt{Manara+2022-PPVII}). It should be kept in mind that our viscous simulations have a $\sigma_{\mathrm{int}}$ of 1 dex for both the disk mass and radius (see Table \ref{tab:parameters}); as large as the intrinsic spread can be, the steeply decreasing viscous trend will always evolve the width of the distribution to $\sigma_{\mathrm{obs}}$. The MHD simulation instead falls within the error bars of the earliest observed star-forming region, up until ages on $\sim 2.5$ Myr. There is an increasing discrepancy for more evolved populations, up until around 20$\%$ for Upper Sco; however, the oldest populations also represent the less complete samples, and therefore they carry a significant bias that should be kept in mind when comparing with simulations. Moreover, there are caveats to our own simulations, as in the viscous case we neglect disk dispersal mechanisms (such as internal or external photoevaporation, e.g. Malanga et al. in prep.) and only consider a detection threshold in disk masses.

\section{Discussion and conclusions}\label{sec:discussion}

In this work, we have investigated how the time evolution of the distribution of a population of disks in the $M_{\mathrm{d}} - \dot M$ plane is impacted by the evolutionary model, considering the viscous and MHD prescriptions respectively. We have presented a combination of analytical considerations and numerical simulations, performed through the 1D population synthesis code \texttt{Diskpop}, in the case of a log-normal distribution of initial accretion timescales (which reproduces both the disk and accretion fraction). We find that, while the mean of the distribution of $t_{\mathrm{lt}} = M_{\mathrm{d}}/\dot M$ is not significantly impacted by the chosen model, the expected behavior of the width shows considerable differences depending on the evolutionary prescription; when including the observational biases in the form of additional uncertainty, this distinctive behavior is maintained.

Our predictions will be exploited to their full potential through a comparison with the results of the current observational effort to obtain direct estimates of disk gas masses; for the time being, we compare our evolutionary trends with the latest available observational data (based on dust observations) in different star-forming regions. We find that the purely viscous case only manages to marginally reproduce the observations at the earliest ages, while the MHD curve resembles them better. Based on these results, we suggest the analysis of these distributions as a viable method to disentangle between the viscous and MHD evolutionary models; our data comparison hints at a better agreement with the MHD model.

\section*{Acknowledgments}

We thank an anonymous referee for their comments that helped us improving the clarity of the manuscript. This work was partly supported by the Italian Ministero dell’Istruzione, Universit\`{a} e Ricerca through the grant Progetti Premiali 2012-iALMA (CUP C52I13000140001), by the Deutsche Forschungsgemeinschaft (DFG, German Research Foundation) - Ref no. 325594231 FOR 2634/2 TE 1024/2-1, by the DFG Cluster of Excellence Origins (www.origins-cluster.de). This project has received funding from the European Union’s Horizon 2020 research and innovation program under the Marie Sklodowska- Curie grant agreement No 823823 (DUSTBUSTERS) and from the European Research Council (ERC) via the ERC Synergy Grant ECOGAL (grant 855130), ERC Starting Grant DiscEvol (grant 101039651) and ERC Starting Grant WANDA (grant 101039452). Views and opinions expressed are however those of the author(s) only and do not necessarily reflect those of the European Union or the European Research Council Executive Agency. Neither the European Union nor the granting authority can be held responsible for them. B.T. acknowledges support from the Programme National ‘Physique et Chimie du Milieu Interstellaire’ (PCMI) of CNRS/INSU with INC/INP and cofunded by CNES.



\bibliography{bibliography}{}
\bibliographystyle{aasjournal}

\appendix

\section{Skewness of the distribution}\label{appendix-skewness}

The skewness of a distribution, defined as the third standardized moment, measures the asymmetry of the distribution about its mean. As we mentioned in paragraph \ref{subsec:lifetimes_distr}, alongside the mean value and the width, in the viscous case we expect also the skewness of the distribution of $t_{\mathrm{lt}}$ to evolve in time; in this Appendix we discuss this theoretical expectation and show the results of our numerical simulations.

\begin{figure}[htbp]
    \centering
    \includegraphics[width = \textwidth]{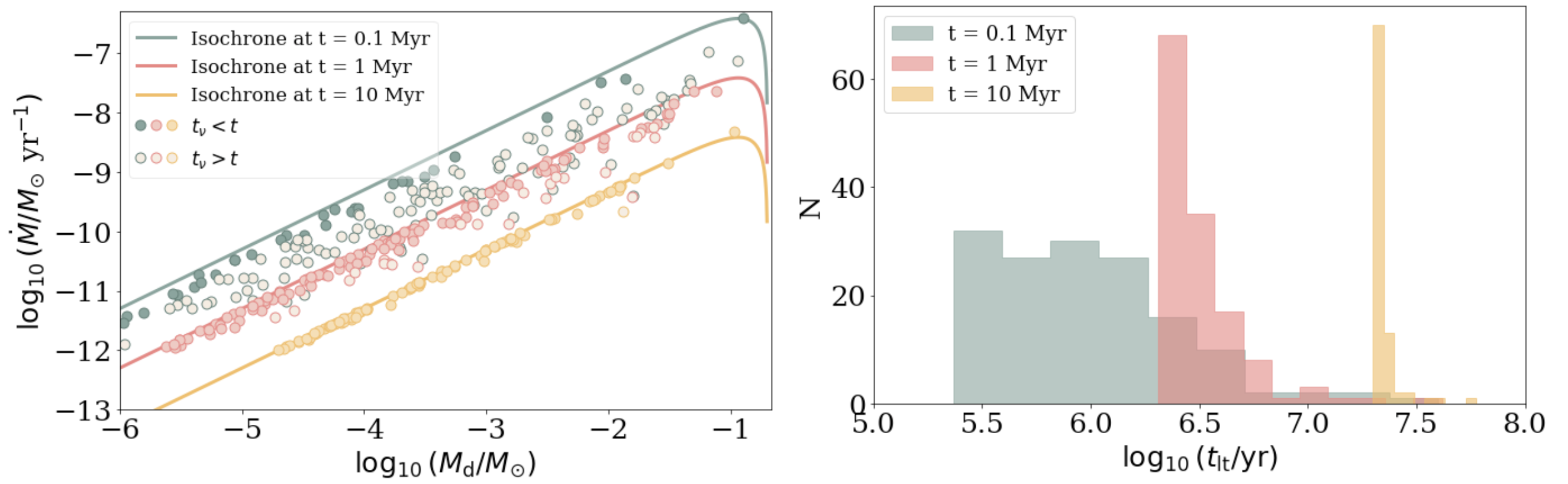}
    \caption{Time evolution of a synthetic population of viscous disks in the $M_{\mathrm{d}} - \dot M$ plane (left panel) and corresponding histograms of $t_{\mathrm{lt}}$ (right panel). The color coding is as in Figure \ref{fig:isochrones_both}. Full dots represent disks whose initial viscous timescale is shorter than the age of the population, and that can therefore be considered evolved.}
    \label{fig:appendix-isochrones}
\end{figure}

\begin{figure}
    \centering
    \includegraphics[width = \textwidth]{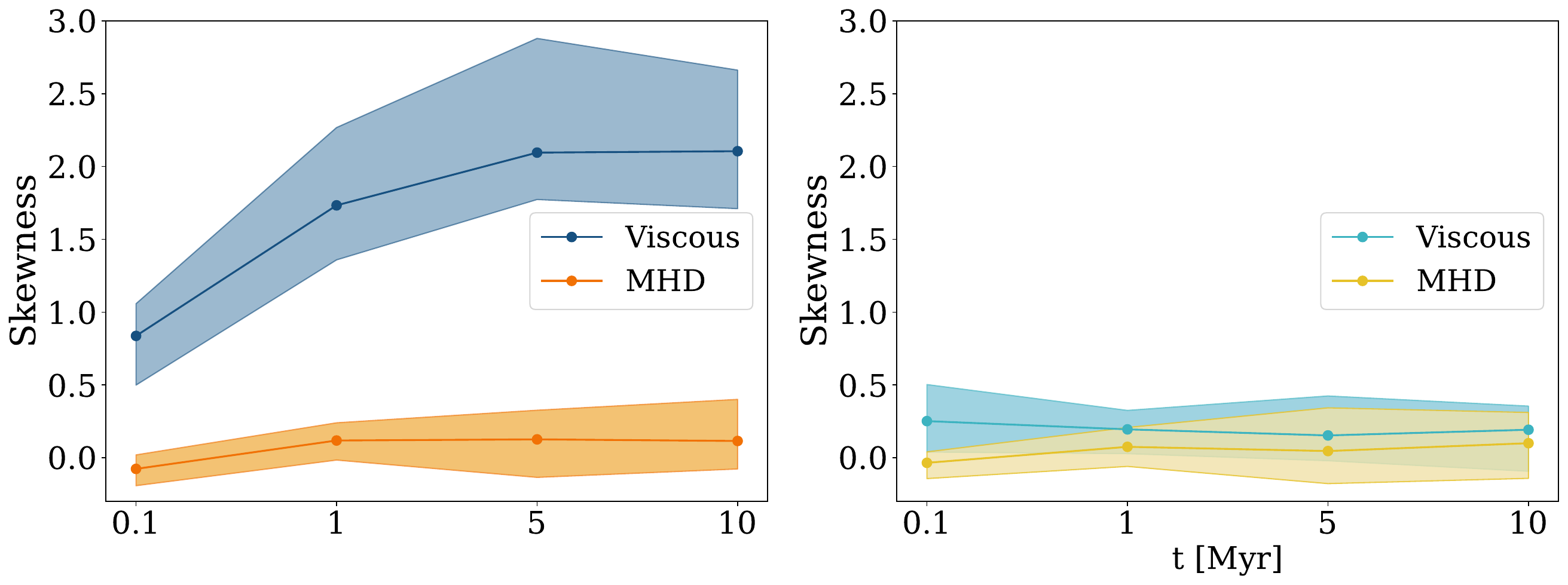}
    \caption{Time evolution of the skewness of the distribution of $t_{\mathrm{lt}}$ for a synthetic population of protoplanetary disks, with the same color coding as Figure \ref{fig:mean_and_width}. Both panels show the comparison between the viscous and MHD models, without (left) and with (right) the additional observational uncertainty $\sigma_{\mathrm{obs}}$. Despite the theoretical predictions of the two models being significantly different (left panel), the convolution with observational biases completely smooths them out (right panel).}
    \label{fig:skewness_all}
\end{figure}

The left panel of Figure \ref{fig:appendix-isochrones} shows a population of viscously evolving disks (dots) at three subsequent ages, as well as the corresponding theoretical isochrones (solid lines). Full dots represent disks whose initial viscous timescale is shorter than the age of the population, which as a whole can therefore be considered evolved: from viscous theory, such disks are expected to have reached the self-similar condition and lie on the analytical isochrone, i.e., to show a linear correlation between the disk mass and the accretion rate. On the other hand, empty dots represent not-yet-evolved disks, which lie below the theoretical isochrone. As the population evolves, more disks satisfy the $t_{\nu} < t$ condition, as can be visualized by the increasing number of full dots in Figure \ref{fig:appendix-isochrones}; this implies that more disks lie on the theoretical isochrone, bringing the population on the $M_{\mathrm{d}} - \dot M$ plane closer to a line. While this causes the width of the distribution of $t_{\mathrm{tl}}$ to decrease with time, the skewness on the other hand increases - as we show in the right panel of Figure \ref{fig:appendix-isochrones}, which represents the corresponding histograms at all ages. This skewing effect is due to the fact that younger disks, which do not lie on the isochrone yet, have a $t_{\mathrm{lt}}$ \textit{longer} than the actual age of the region, and therefore contribute to positively skew the distribution - while evolved disks, which make up the bulk of the population, cluster close to the mean value. Figure \ref{fig:skewness_all} shows the evolution of the skewness of a population of disks generated and evolved with \texttt{Diskpop} with the same color coding and shaded areas as Figure \ref{fig:isochrones_both}; the left panel represents the case with no observational uncertainty, where the viscous distribution (blue) gets more and more skewed as expected, growing by a factor of 2 between 0.1 and 10 Myr. On the other hand, the MHD distribution (orange) remains symmetrical within the 25th-75th percentile for the whole evolution, resulting in a factor 3 difference from the viscous model for evolved populations. As significant as this theoretical difference is, including the observational biases (right panel) completely smooths it out: the two expected observed behaviors are indistinguishable once convoluted with the additional observational uncertainties. 

In conclusion, while the evolution of the skewness makes an interesting theoretical argument stemming from the different interpretation of isochrones in the two models, it does not provide a reliable method to compare viscosity and MHD from the observational point of view.

\section{Time evolution of the distribution of \titlelowercase{\texorpdfstring{$t_{\mathrm{lt}}$}{tlt}}}\label{appendix:tlt}

As $t_{\mathrm{lt}}$ depends on $t_{\mathrm{acc}, 0}$ as $t_{\mathrm{lt}} = (1 + f_M)(2t_{\mathrm{acc}, 0} - \omega t)$, the evolved distribution of $t_{\mathrm{lt}}$ is determined by the choice of initial distribution of $t_{\mathrm{acc}, 0}$: \cite{Tabone+2021b} have shown that, choosing an exponential distribution for $t_{\mathrm{acc}, 0}$, the corresponding distribution of $t_{\mathrm{lt}}$ reads

\begin{equation}
    \frac{dP}{d t_{\mathrm{lt}}} = \frac{1}{\omega \tau (1 + f_M)} \exp{\left(- \frac{t_{\mathrm{lt}}}{(1+f_M) \omega \tau}\right)} f_D(t),
    \label{eq:distr_tlt_tabone}
\end{equation}

where $f_M$ is defined in \cite{Tabone+2021a} and $\tau = 2.5$ Myr to fit the disk fraction, $f_D(t) = \exp{(-t/\tau)}$. As $f_D$ is only a normalization factor, \eqref{eq:distr_tlt_tabone} still have an exponential shape; moreover, it does not depend on time, as well as its mean value. On the other hand, if we pick a log-normal distribution for  $t_{\mathrm{acc}, 0}$, we can still reproduce both the disk and the accretion fraction (see Appendix \ref{appendix:df_and_af}) but in that case the evolved distribution of $t_{\mathrm{lt}}$ becomes

\begin{equation}
    \label{eq:distr_tlt_lognormal}
    \begin{split}   
        \frac{dP}{dt_{\mathrm{lt}}} = &  \frac{1}{\sqrt{2 \pi \sigma^2}} \frac{1}{t_{\mathrm{lt}} + (1+f_M) \omega t} \exp{ \Biggl\{ -\frac{1}{2 \sigma^2} \left[ \log{\left( \frac{t_{\mathrm{lt}}}{2(1+f_{M})} + \frac{\omega t}{2} \right) - \mu} \right]^2 \Biggl\} },
    \end{split}
\end{equation}

\noindent where $\mu$ and $\sigma$ are the mean value and width of the initial log-normal distribution. Notice that Equation \eqref{eq:distr_tlt_lognormal} is not a log-normal in $t_{\mathrm{lt}}$; moreover, it does depend on time, and so does its mean value and spread.

\begin{figure}
    \centering
    \includegraphics[width = \textwidth]{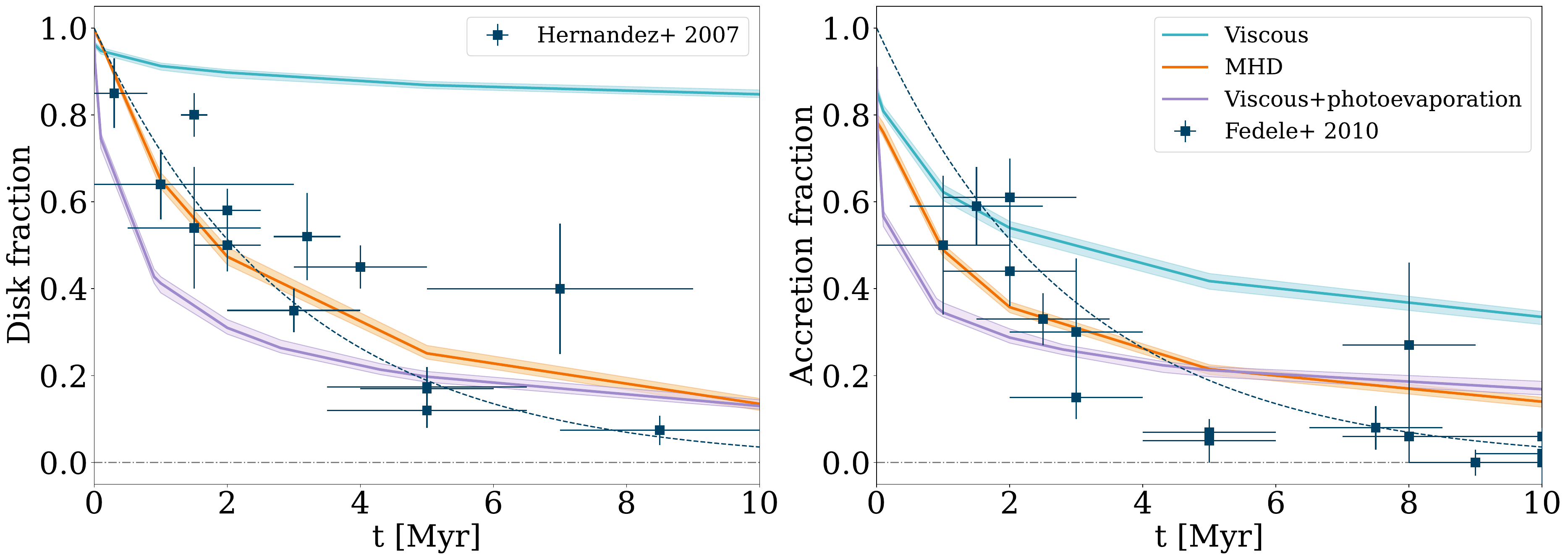}
    \caption{Disk and accretion fraction (left and right panel respectively) in our viscous (light blue), MHD (orange) and viscous+photoevaporation (lilac) simulations, compared with data by \cite{Hernandez+2007} and \cite{Fedele+2010} (blue squares). The shaded areas are as in Figure \ref{fig:mean_and_width}. The dashed blue lines show the exponential fits to the data. Following the original paper, we define the accretion fraction as the fraction of sources with accretion rate higher than $10^{-11}$ M$_{\odot}$/yr. Our choice of a log-normal distribution of initial accretion timescales for the MHD model reproduces both the disk and accretion fraction, as does the exponential distribution chosen by \cite{Tabone+2021b}. The viscous model does not reproduce any of the fractions due to the lack of a disk dispersal mechanism, while including internal photoevaporation allows to recovered the observed behavior.}
    \label{fig:appendix_DF}
\end{figure}

\section{Impact of internal photoevaporation}\label{appendix:df_and_af}

As mentioned in the main paper, disk dispersal in an intrinsic feature of MHD winds. These models manage to reproduce both the disk and accretion fraction, defined as the fraction of young stars with infrared excess \citep{Hernandez+2007} and accreting (i.e., with $\dot M > 10^{-11}$ M$_{\odot}$ yr$^{-1}$ following \citealt{Fedele+2010}) objects respectively, as shown by the orange lines in Figure \ref{fig:appendix_DF}. On the other hand, purely viscous models do not account for disk dispersal. This leads to a mismatch between the predicted and observed disk and accretion fraction, represented by the blue lines in Figure \ref{fig:appendix_DF}: the disk fraction is almost constant to 1, the little decrease being due to the observational threshold that we introduced in our simulations (considering dispersed disks with masses lower than $10^{-6}$ M$_{\odot}$, see Section \ref{subsec:popsynth}), while the accretion fraction does decrease, but not enough to match the observed values. This problem is usually overcome in the literature by including internal photoevaporation, a two-timescale process that introduces a disk dispersal mechanism, allowing to reproduce the observations as shown by the purple lines in Figure \ref{fig:appendix_DF}. We ran the test simulation presented in this Appendix using the standard photoevaporative model of \cite{Owen+2012}, with a mass-loss rate of $10^{-1-}$ M$_{\odot}$ yr$^{-1}$, consistent with the latest constraints \citep{Alexander+2023}.

Once internal photoevaporation kicks in, it lowers the accretion rates for a given disc mass, introducing therefore a spread in the $M_{\mathrm{d}} - \dot M$ plane \citep{Somigliana+2020}; therefore, it could in principle affect the conclusions of this work. However, we have tested that the mean and width of the $t_{\mathrm{lt}}$ distribution in the presence of photoevaporation do not significantly deviate from the purely viscous prediction; without observational spread the photoevaporative case lies between the viscous and MHD models, and becomes indistinguishable from the viscosity when the observational spread is included.

\label{lastpage}

\end{document}